\documentclass[onecolumn, showpacs, pre]{revtex4}

\usepackage{graphicx}
\usepackage{verbatim}
\DeclareGraphicsExtensions{.eps}

\usepackage{amsmath}
\usepackage{amsfonts}
\usepackage{ctable}
\usepackage{rotating}
\usepackage{multirow}
\usepackage{subfig}
\begin{document}

\title{Degree Correlations in Random Geometric Graphs
}


\author{A. Antonioni} \email[]{alberto.antonioni@unil.ch} 
\author{M. Tomassini
} \email[]{marco.tomassini@unil.ch} 


\affiliation{Information Systems Department, \\
Faculty of Business and Economics,\\
University of Lausanne, Switzerland
}

\date[]{Received: date / Accepted: date}

\begin{abstract}
Spatially embedded networks are important in several disciplines. The prototypical spatial network we assume is
the Random Geometric Graph of which many properties are known. Here we present new results for the
two-point degree correlation function in terms of the clustering coefficient of the graphs for two-dimensional
space in particular, with extensions to arbitrary finite dimension.

\end{abstract}

\pacs{02.10.Ox, 89.75.-k}
	
\maketitle

In the last decade, thanks to abundant data, new models, and adequate software tools, complex networks have been 
thoroughly investigated in many disciplines and as
a substrate of many phenomena. A synthesis is now emerging, as can be seen e.g. in the recent comprehensive
treatment by Newman~\cite{newman-book} or in Boccaletti et al.~\cite{bocca}.
Most of this work has dealt with ``relational'' networks, i.e. graphs in which 
distances do not have physical meaning and are just dimensionless quantities measured in terms of 
edge hops. Indeed,  many networks are mainly of this kind such as big social networks. However, in many cases 
the physical space in which networks are 
embedded and the actual distances between nodes are important such as in rail and road networks, ad hoc communication
device networks, and other
geographical and transportation networks.
The recent comprehensive review by Barth\'elemy~\cite{SpatNets} has at last put together a large amount of scattered
material on spatial networks.  The Random Geometric Graph  (RGG) is a standard spatial network model that plays a role for 
spatial networks similar to the one played by  the Erd\"os-R\'enyi random graph for relational ones. This model
is well known~\cite{Penrose,dall02,SpatNets} but some of its second-order features have not yet been uncovered. Among these, there
is the question of the degree correlation functions. In this work we present some results on degree correlations
on RGGs that we believe were previously unknown.

The construction process of a RGG with $N$ nodes and radius $R$ can be summarized as follows~\cite{Penrose,dall02}:
\begin{itemize}
\item the $N$ nodes are placed on the unitary space $\Omega\in\mathbb{R}^d$ with uniform distribution. 
\item an edge is created for every pair of nodes whose distance is $r < R$. 
\end{itemize} 
The distance is given by some metric on $\Omega$. In this work we have dealt with 2-dimensional RGGs and the Euclidean metric distance on $\mathbb{R}^2$. The unitary space $\Omega$ is the square $[0,1]^2$  with no boundary conditions (torus).



It is also possible to adopt different shapes of neighborhood area generated according to other metrics. For example, the Manhattan distance is sometimes used to model mobility networks~\cite{Glauche2003577}. We found that the general properties of these networks are very close to those using the more common Euclidean distance, which are the ones we describe here.

The average degree $\overline k=\overline k(N,R)$ of a RGG can be easily estimated by the formula $\overline k=\rho V$, where $\rho$ is the node density, representing the number of nodes within a unit space, and $V$ is the neighborhood area volume. In this case $\rho=N$, since $\Omega$ is an unitary space, and $V=\pi R^2$. In conclusion, $\overline k=\pi NR^2$. According to this result, it is possible to consider $\overline k$ as a parameter of RGGs, instead of the radius $R$. Therefore, in order to construct RGG with an average degree that tends to $\overline k$ as $N,1/R\rightarrow\infty$, it is sufficient to use the radius $R=\sqrt{\bar k/(\pi N)}$. 

The degree distribution of RGGs can be estimated regarding the probability density function of having a node $X$ of degree $k$, given that there are other $N$ nodes uniformly distributed in $\Omega$. More precisely, $N-1$ other nodes, but $N\sim N-1$ for large values of $N$. 
This probability follows the binomial distribution and it is equal to:
$$\mathbb{P}(k_X=k)=\binom{N}{k}p^{k}(1-p)^{n-k}$$
where $p=\pi R^2$, since it represents the proportion between V and $\Omega$. 
The Poisson distribution with parameter $\lambda=Np$ can be used as an approximation of the binomial distribution if $N$ is sufficiently large and $p$ is sufficiently small. In this case the degree distribution will be approximated by: 
$$\mathbb{P}(k_X=k)=\frac{\lambda^{k}}{k!}e^{-k}$$
where $\lambda=\overline k$. 
%

The average clustering coefficient is given averaging on all node's individual clustering 
coefficients~\cite{newman-book,bocca,watts-strogatz-98}. This property on RGGs was extensively studied in the work of Dall and Christensen \cite{dall02}, in which they have found the law for the average clustering coefficient as a function of the dimension of the space. Here the dimension is equal to $2$, and it is possible to demonstrate that the average clustering coefficient $c_2$ tends to $A_2=1-\frac{3\sqrt 3}{4\pi}\sim0.5865$, for large values of $N$ and for all 2-dimensional RGGs~\cite{dall02} in the 
Euclidean space. By analogy, we shall call $c_d$  the average clustering coefficient of a $d$-dimensional RGG. \\
This important result depends on the particular construction of RGGs. The average clustering coefficient tends to the ratio of the average shared neighborhood area of two connected nodes and the whole neighborhood area. It is clear that changing the radius $R$ this fraction maintains the same value. This phenomenon, which conducts to a fixed average clustering coefficient for every RGGs, will be studied in depth in order to estimate the degree correlations in the following. Many other properties of RGGs
have been studied in Penrose's book~\cite{Penrose}.

\begin{figure}[ht]\centering
\includegraphics[width=0.45\textwidth]{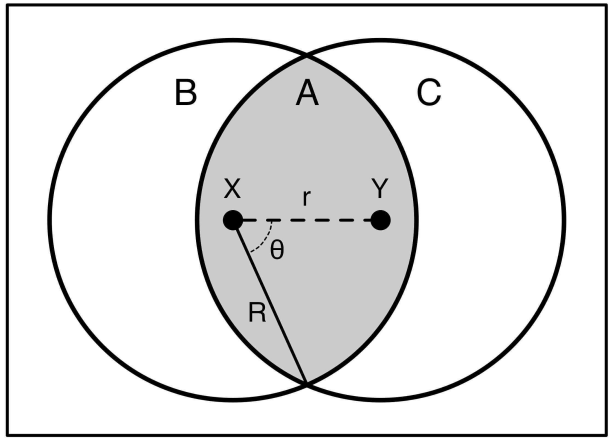}
\caption{Two connected nodes $X,Y$ and their neighbor\-hood areas of ra\-dius $R$. The dis\-tance be\-tween them is $r$ and the angle is $\theta=\arccos(\frac{r}{2R})$. The grey area $A$ is the shared area of the two connected nodes. $B$ and $C$ are the complementary areas of $A$ to $V$.}
\label{area}
\end{figure}

Several studies were conducted on the relationship between degree-degree correlations and clustering coefficients in relational networks~\cite{boguna,dorogo,pusch}. Here, we focus on the direct dependence between the clustering coefficient and degree-degree correlation in RGGs.
Due to its construction process, in a RGG there is positive degree-degree correlation. This property is commonly detected studying the average degree of the neighborhood of a given node of degree $k$~\cite{satorras-corr}. The function $k_{nn}(k)$ (\emph{nearest neighbor average degree}) represents the average degree of the neighborhood of all nodes of degree $k$. The properties which emerge from the spatial construction of RGGs allow us to evaluate $k_{nn}(k)$ with a mean-field method for very large values of $N$. 
It is possible to find the average degree of neighbors node estimating the average shared area of two connected nodes. Fig. \ref{area} depicts the case of two connected nodes $X$ and $Y$, and their shared area $A$ (grey). $B$ is the complementary area of $A$ to neighborhood area $V=\pi R^2$, and $r$ is the distance between the two nodes. The area $C$ is symmetrical and equivalent to $B$.

The shared area $A(r)$ of two connected nodes is only dependent on the distance $r$ between them. The formula for $A(r)$ can be simply derived from those of the circular sector and is equal to:
\begin{equation}
A(r)=R^2(2\theta-\sin2\theta),\;\;\;\;\text{where } \theta=\arccos\Big(\frac{r}{2R}\Big)
\end{equation}
Thus, the average shared area $\bar A$ is obtained integrating and averaging $A(r)$ on all possible neighbors $Y$ of $X$ (see Fig.~\ref{area}). 
We now calculate this area:
\begin{equation}
\bar A =\bar A(R) = \frac{\int_0^{2\pi}\int_0^R R^2(2\theta-\sin2\theta)rdrd\phi}{\pi R^2}
\label{aver}
\end{equation}
where variables $0\leq r\leq R$ and $0\leq\phi\leq 2\pi$ represent all possible neighbors $Y$ in the neighborhood area of $X$. \\
The numerator in (\ref{aver}) is calculated by using the substitutions $r=2R\cos\theta$ and $\theta=\psi/2$, which leads to:
\begin{equation}
-R^4\int_0^{2\pi}\int_{\pi}^{\frac{2}{3}\pi} (\psi-\sin\psi)\sin\psi\;d\psi d\phi=R^4\Big(\pi^2-\frac{3\sqrt 3}{4}\pi\Big)
\label{aver2}
\end{equation}
From (\ref{aver}) and (\ref{aver2}), it follows that:
\begin{equation}
\bar A = R^2\Big(\pi-\frac{3\sqrt3}{4}\Big)
\label{aver3}
\end{equation}
According to (\ref{aver3}), it is possible to evaluate the ratio of $\bar A$ and $V$, which leads to:
\begin{equation}
\frac{\bar A}{V}=\frac{\bar A}{\pi R^2}=1-\frac{3\sqrt 3}{4\pi}=A_2
\end{equation}
where $A_2\sim 0.5865$ is the asymptotic value of the average clustering coefficient of 2-dimensional RGGs. \\
Finally, using the mean-field method to evaluate the neighbors shared area, it is possible to find the expression for the function $k_{nn}(k)$. 
In order to understand better this result, it is useful to use again the notation of Fig.~\ref{area}. Focusing on node $X$ of known degree $k_X$, we are trying to evaluate the average degree of its neighbor $Y$. This degree is given by two different areas, where the nodes density could be different. \\
The first area which brings neighbors to node $Y$ is the shared area $A$, which is approximated by $\bar A$ in the mean-field method, and where the nodes density is equal to $\frac{k_X}{V}$, thanks to the fact that we know the degree of $X$. \\
The second area in which node $Y$ could find other neighbors is the complement $C$ of $A$. In particular, approximating $A$ 
by $\bar A$ we approximate also $C$ to $\overline C=V-\bar A$. In $C$ the nodes density is equal to $\rho=N$, since we
do not have more information about this area. \\
Putting together this information we can easily give the expression for $k_{nn}(k_X)$, which represents the degree of node $Y$:
\begin{equation}
k_{nn}(k_X)=\frac{k_X}{V}\bar A+\rho(V-\bar A)=A_2\cdot k_X+(1-A_2)\overline k
\end{equation}
But $k_X$ stands for the degree of a generic node in the graph and this leads to the final expression of the function:
\begin{equation}
k_{nn}(k)=A_2\cdot k+(1-A_2)\overline k
\end{equation}
This function is linear in $k$ and reveals the positive relation between the degree of a node and its average neighbors degree ($A_2>0$). 

\begin{figure}[htb]
\begin{center}
\includegraphics[height=0.35\textwidth,width=0.45\textwidth]{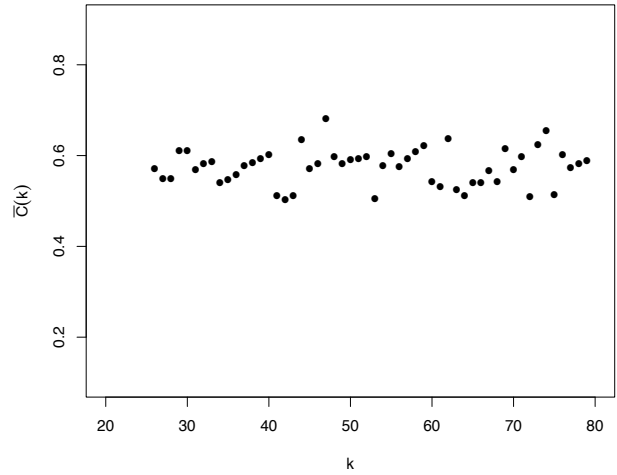}
\caption{Mean clustering of a node as a function of the degree of that node in an instance of a RGG. $N=50000$, $\bar k=50$.}
\label{cc-node}
\end{center}
\end{figure}

Another interesting quantity is the mean clustering coefficient as a function of the node degree $\bar C(k)$, whose empirical
values for an instance of a RGG are plotted in Fig.~\ref{cc-node}. From the figure it clearly appears that this measure is
independent of $k$ and that its mean value tends to the average clustering coefficient $A_2$ as computed above. This is
similar to what happens in Erd\"os-R\'enyi random graphs in which the average clustering is constant and equal to
the probability $p$ of existence of an edge~\cite{newman-book}. Here the role of $p$ is played by the ratio $\bar A/V$ of the areas as explained above.

\begin{figure*}[ht]
\begin{center}
\subfloat[]{\label{k50}\includegraphics[width=0.45\textwidth]{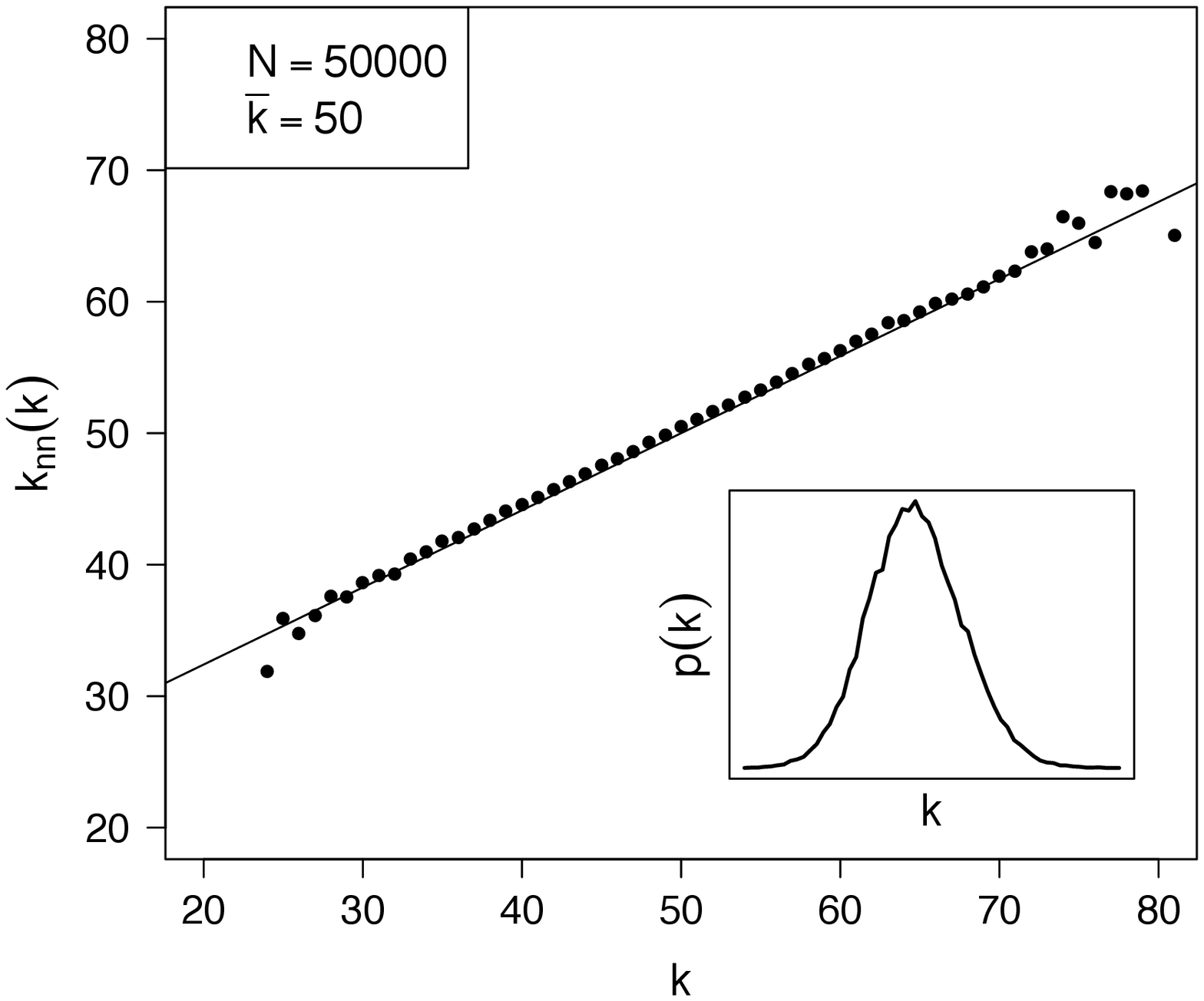}}
\hspace{0.6cm}
\subfloat[]{\label{k100}\includegraphics[width=0.45\textwidth]{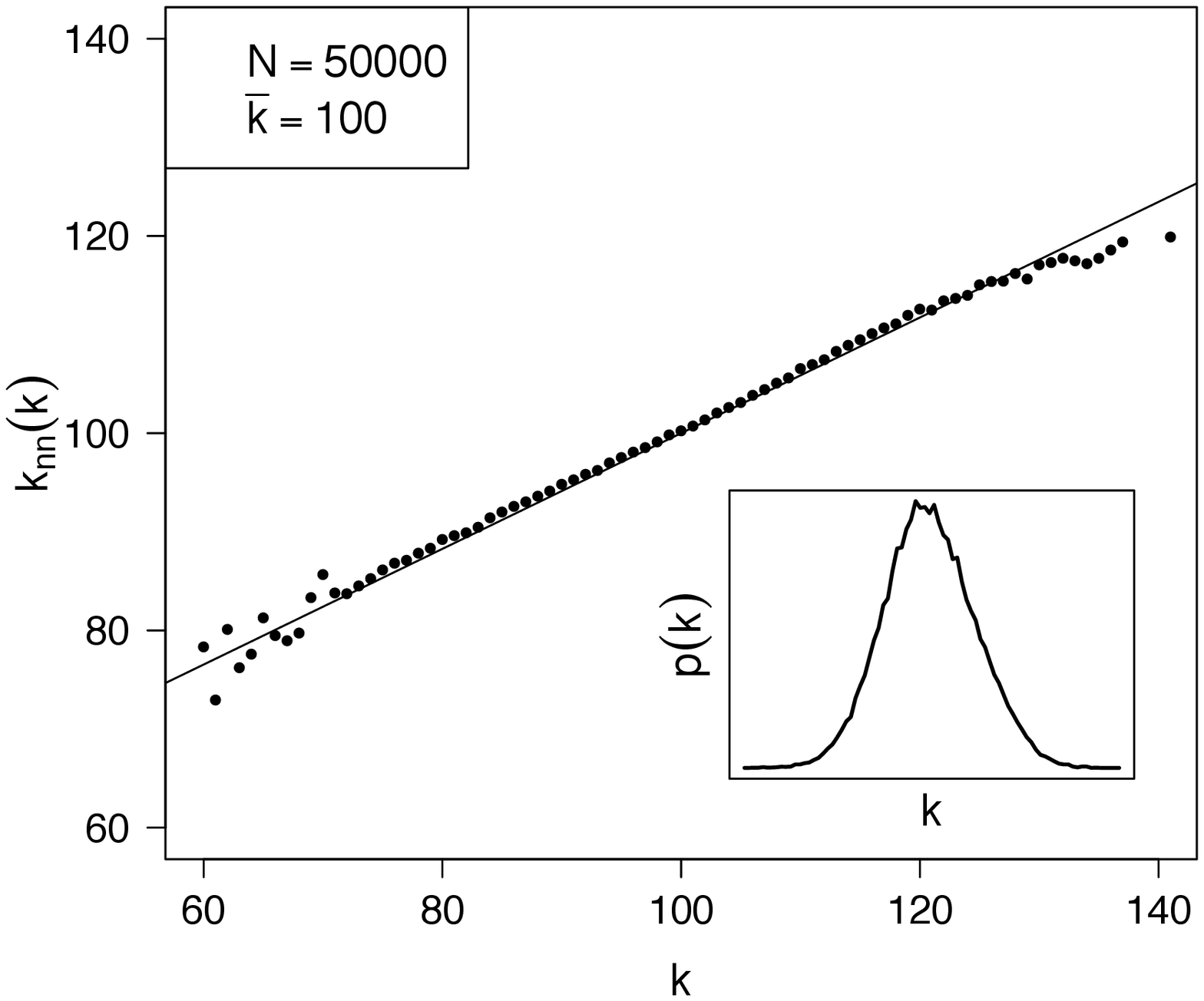}}
\caption{Straight thick line: theoretical $k_{nn}(k)$ with a slope coefficient of $0.5865$. Dotted curve: empirical $k_{nn}(k)$.
(a): $N=50000$, $\overline k= 50$. (b): $N=50000$, $\overline k= 100$. Due to the peaked empirical degree distribution functions (see insets) low and high $k$ values are noisier.}
\label{asso1}
\end{center}
\end{figure*}

The degree-degree correlation is not only estimated by $k_{nn}(k)$, but it is also exactly given by the \emph{assortativity coefficient}, which is the Pearson correlation coefficient of degree between pairs of connected nodes~\cite{newman-02,bocca}. This value is widely used as a measure of the strength of linear dependence between two variables \cite{newman-book}. As for the average clustering coefficient, we use the notation $r_2$ to indicate the assortativity coefficient of a 2-dimensional RGG. \\
As we have seen above, the function $k_{nn}(k)$ is linear and, applying the mean-field method, it can represent the regression line for the two variables, degree ($X$) and neighbor degree ($Y$). We thus assume that $k_{nn}(k)$ is the regression line and we 
derive $r_2$ from that. \\
The regression line slope $b$ tends to $A_2$ for large values of $N$ and is defined by the following formula:
\begin{equation}
b=\frac{cov(X,Y)}{\sigma_X^2}\rightarrow A_2,\;\; N \rightarrow \infty
\end{equation}
where $\sigma_X$ is the standard deviation of variable $X$ and $cov(X,Y)$ is the covariance of the two variables $X,Y$. \\
On the other hand, $r_2$ by definition is the covariance of the two variables $X,Y$ divided by the product of their standard deviations:
\begin{equation}
r_2=\frac{cov(X,Y)}{\sigma_X\sigma_Y}
\end{equation}
It follows this last result:
\begin{equation}
r_2\rightarrow\frac{\sigma_X}{\sigma_Y}A_2
\end{equation}
where $\sigma_X, \sigma_Y$ are the standard deviations of variables $X$ and $Y$, respectively. \\
In order to estimate these two standard deviations, we focus on their distribution functions $f_X(x)$ and $f_Y(y)$. 
We already know that $f_X(x)$ is a Poisson distribution since it represents the degree distribution of a RGG. This implies that $\sigma_X=\overline k$. \\
However, we can find the expression of distribution $f_Y(y)$, which is the distribution of neighbors degrees, using the relation:
\begin{equation}
f_Y(y)=\frac{f_{Y}(y|x=x^*)}{f_{X}(x|y=y^*)}f_X(x)
\label{func}
\end{equation}
The numerator function $f_Y(y|x=x^*)$ is the degree distribution of neighbors of a given node of degree $x^*$, and which has an expected value equal to $k_{nn}(x^*)$. \\
The other function $f_{X}(x|y=y^*)$ represents the opposite case. This function is the probability distribution of nodes degrees given that they have a neighbor of degree equal to $y^*$. \\
Since the two functions are completely equivalent because of the symmetry of their definitions, we can conclude from~(\ref{func}) that $f_Y(y)=f_X(x)$, and, consequently, $\sigma_X=\sigma_Y$. We can then conclude that $r_2\rightarrow A_2$ for large values of $N$. \\
In Fig. \ref{asso1} we depict  $k_{nn}(k)$, theoretical and empirical, in two RGGs with $N=50000$ and $\overline k = 50, 100$. From
the figures, one can conclude that there is a very good agreement with the theoretical results. 

This last result can be extended to other kinds of RGGs, with different dimensions or neighborhood volume (for $d \ge 3$), since it does not depend on the shapes of the neighborhood volume, but only on the ratio of the average shared volume and the neighborhood volume. For the RGGs this ratio is intrinsically represented by the average clustering coefficient of the graph. The individual clustering coefficient $C_X$ \cite{newman-book,newman-03} of a node $X$ is given by the following definition:
$$T_{XY}=|\{\text{triangles with edge } (X,Y)\}|$$
which represents the number of all triangles formed with the edge $(X,Y)$ in the graph. It follows that:
\begin{equation}
C_X=\frac{1}{k_X(k_X-1)}\sum_{Y\in V_X} T_{XY}
\label{Cx}
\end{equation}
where $V_X$ is the set of the neighbors of $X$. \\
Let us denote $A_d$ the ratio of the average shared volume of two connected nodes and the neighborhood volume
 of $d$-dimensional RGGs. We find that, on average, $T_{XY}\rightarrow~(k_X-1)A_d$ and, substituting in~(\ref{Cx}), that $C_X\rightarrow A_d$. Thus, the average clustering coefficient $c_d = \frac {1} {N} \sum_X C_X$ of $d$-dimensional RGGs tends to $A_d$. \\
Now, considering the formula which estimates $c_d$ in the Euclidean space by Dall and Christensen~\cite{dall02}:
\begin{equation}
c_d\sim 3\sqrt{\frac{2}{\pi d}}\Big(\frac{3}{4}\Big)^\frac{d+1}{2}
\label{cd}
\end{equation}
we can conclude that~(\ref{cd}) represents a good approximation of $A_d$ for large values of $N$ and $d$, while, for small values of $d$, this function overestimates $A_d$. \\
The constant $A_d$ depends on the neighborhood volume shape and represents the asymptotic value for the average clustering coefficient $c_d$ and the assortativity coefficient $r_d$. The last assertion is due to the fact that, as we have seen above in the case $d=2$, $r_2$ tends to the fraction $A_2$. This process is applicable to any dimension $d$ in order to evaluate $r_d$.

\begin{figure}[htb]
\begin{center}
\includegraphics[height=0.35\textwidth,width=0.45\textwidth]{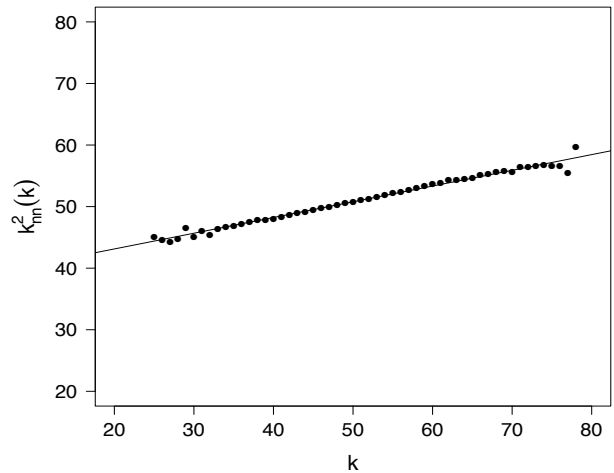}
\caption{Dotted curve: empirical $k^{2}_{nn}(k)$. $N=50000$, $\overline k= 50$. Straight thick line: linear regression line
with slope coefficient of $0.255$. }
\label{asso2}
\end{center}
\end{figure}

Similar analytical results can be obtained in the same way extending to higher order of degree correlations, but the amount of calculus becomes particularly heavy. For example, the correlation coefficient between a given node's degree and the 
degree of its neighbors at distance $2$ can be obtained from
the study of the function $k_{nn}^2(k)$, which represents the average degree of neighbors at distance $2$. Here the
distance is intended to be the relational distance in the graph, i.e. the number of edges that compose the minimum shortest path which connects the two nodes~\cite{sinatra}. We thus calculated numerically $k_{nn}^2(k)$ for an instance of a RGG, plotted 
the results, and computed the regression line with standard tools as shown in Fig.~\ref{asso2}. \\
From Figs.~\ref{asso1} and \ref{asso2} one sees that the degree correlations are non-negligible up to graph distance equal to $2$.
However, they decrease going from distance one to two and, given the way in which the RGG is built,  we hypothesize that they 
tend to vanish for larger distances. 

In summary, we have presented new results for the degree correlations in RGGs, showing exact results for the two-dimensional case and extending them
to arbitrary finite dimension.

\bibliographystyle{unsrt}
\bibliography{rgg}

\begin{thebibliography}{10}

\bibitem{newman-book}
M.~E.~J. Newman.
\newblock {\em Networks: An Introduction}.
\newblock Oxford University Press, Oxford, UK, 2010.

\bibitem{bocca}
S.~Boccaletti, V.~Latora, Y.~Moreno, M.~Chavez, and D.-U. Hwang.
\newblock Complex networks: Structure and dynamics.
\newblock {\em Physics Reports}, 424(4‰ÛÒ5):175 -- 308, 2006.

\bibitem{SpatNets}
M.~Barth\'elemy.
\newblock Spatial networks.
\newblock {\em Physics Reports}, 499:1--101, 2011.

\bibitem{Penrose}
M.~Penrose.
\newblock {\em Random Geometric Graphs}.
\newblock Oxford University Press, Oxford, UK, 2003.

\bibitem{dall02}
J.~Dall and M.~Christensen.
\newblock Random geometric graphs.
\newblock {\em Phys. Rev. E}, 66:016121, 2002.

\bibitem{Glauche2003577}
I.~Glauche, W.~Krause, R.~Sollacher, and M.~Greiner.
\newblock Continuum percolation of wireless ad hoc communication networks.
\newblock {\em Physica A: Statistical Mechanics and its Applications}, 325:577
  -- 600, 2003.

\bibitem{watts-strogatz-98}
D.~J. Watts and S.~H. Strogatz.
\newblock Collective dynamics of `small-world' networks.
\newblock {\em Nature}, 393:440--442, 1998.

\bibitem{boguna}
M.~Bogu\~n\'a and R.~Pastor-Satorras.
\newblock Class of correlated random networks with hidden variables.
\newblock {\em Phys. Rev. E}, 68:036112, 2003.

\bibitem{dorogo}
S.~N. Dorogovtsev.
\newblock Clustering of correlated networks.
\newblock {\em Phys. Rev. E}, 69:027104, 2004.

\bibitem{pusch}
A.~Pusch, S.~Weber, and M.~Porto.
\newblock Impact of topology on the dynamical organization of cooperation in
  the prisoner's dilemma game.
\newblock {\em Phys. Rev. E}, 77:036120, 2008.

\bibitem{satorras-corr}
R.~Pastor-Satorras, A.~V\'azquez, and A.~Vespignani.
\newblock Dynamical and correlation properties of the {I}nternet.
\newblock {\em Phys. Rev. Lett.}, 87:258701, 2001.

\bibitem{newman-02}
M.~E.~J. Newman.
\newblock Assortative mixing in networks.
\newblock {\em Phys. Rev. Lett.}, 89:208701, 2002.

\bibitem{newman-03}
M.~E.~J. Newman.
\newblock The structure and function of complex networks.
\newblock {\em {SIAM} Review}, 45:167--256, 2003.

\bibitem{sinatra}
R.~Sinatra, J.~G\'omez-Garde\~nes, R.~Lambiotte, V.~Nicosia, and V.~Latora.
\newblock Maximal-entropy random walks in complex networks with limited
  information.
\newblock {\em Phys. Rev. E}, 83:030103, 2011.

\end{thebibliography}

\end{document}